\documentclass[12pt,reqno,a4paper]{amsart}
\usepackage{amssymb,latexsym}
\usepackage{amsthm, amsmath, amsfonts}
\usepackage{hyperref}
\usepackage{empheq}
\usepackage{graphicx}  
\usepackage{dcolumn}   
\usepackage{bm}        
\usepackage{empheq}
\usepackage{enumerate}
\usepackage{slashed}
\usepackage{simplewick}
\usepackage{comment}
\usepackage{color}
\usepackage{soul}
\usepackage[english]{babel}
\usepackage{appendix}
\usepackage[utf8]{inputenc}
\usepackage{tikz}

\usepackage{marginnote}
\usepackage[normalem]{ulem}

\def\be{\begin{equation}}
\def\ee{\end{equation}}
\def\bea{\begin{eqnarray}}
\def\eea{\end{eqnarray}}
\def\bz{\bar z}

\def\bs{\bar s}
\def\p{\partial}
\def\bp{\bar\partial}

\def\bs{\bar s}

\def\bcal_k{\mathcal B_k}

\newcommand{\kahler}{K\"ahler }

\newtheoremstyle{dotless}{}{}{\itshape}{}{\bfseries}{}{ }{}
\theoremstyle{dotless}

\begin{document}

\title[Lowest Landau level on a cone and zeta determinants]
{Lowest Landau level on a cone\\ and zeta determinants}
\author[Semyon Klevtsov]{Semyon Klevtsov}

\maketitle

\begin{center}
{\small
\address{\it Institut f\"ur Theoretische Physik, Universit{\"a}t zu K{\"o}ln,\\ Z\"ulpicher Str.\ 77, 50937 K{\"o}ln, Germany}}

\end{center}

\vspace{.5cm}
\begin{abstract} 
We consider the integer QH state on Riemann surfaces with conical singularities, with the main objective of detecting the effect of the gravitational anomaly directly from the form of the wave function on a singular geometry. We suggest the formula expressing the normalisation factor of the holomorphic state in terms of the regularized zeta determinant on conical surfaces and check this relation for some model geometries. We also comment on possible extensions of this result to the fractional QH states. 
\end{abstract}
\vspace{1cm}

\thispagestyle{empty}

\section{Introduction}

Considering quantum Hall states on curved backgrounds can help elucidate novel transport coefficients, hidden in the planar geometry. As an example, the central charge for the trial quantum Hall states, arising due to the gravitational anomaly, can be determined this way \cite{K,AG1,CLW,AG2,FK,AG4,CLW1,LCW,BR1,KW}. The gravitational anomaly in QH states appears in the large $N$ expansion of the generating functional $\log Z$, where $Z$ is the normalization factor of the holomorphic QH wave function. For the integer QH state and for the Laughlin state $Z$ has another interpretation as the partition function of the 2d Coulomb plasma. On a compact Riemann surface $\Sigma$ in a constant perpendicular magnetic field with total flux $k$  the generating functional admits the following asymptotic expansion at large $k$,
\begin{equation}\label{logZintr}
\log Z[g_0,g]=\sum_{m=0}^{\infty}k^{2-m}c_{2-m}S_{2-m}(g_0,g).
\end{equation}
Here $g_0$ is the background metric on $\Sigma$, which can be taken as constant scalar curvature metric, $g$ is an arbitrary curved metric on $\Sigma$ and $S_{m}$ denotes certain geometric functionals, see Eqns.\ (\ref{SAY})-(\ref{SL}). Since the number of particles $N$ and the total flux of the magnetic field $k$ are related by the Riemann-Roch formula $N=k+1-{\rm g}$, where ${\rm g}$ is the genus of $\Sigma$, large $N$ and large $k$ expansions are essentially equivalent. Eq.\ \eqref{logZintr} was derived by various methods for 2d Coulomb plasma in the planar domains \cite{ZW}, for the integer QH state on compact Riemann surfaces \cite{K,KMMW} and for the fractional QH states on Riemann surfaces \cite{CLW,FK,CLW1,LCW}. 

The gravitational anomaly appears at the $O(1)$ order in the expansion above in the form of the Liouville functional $S_0(g_0,g)=S_L(g_0,g)$, see Eq.\ \eqref{SL}. Recall the formula for the conformal variation of the regularized determinant of spectral laplacian
\begin{equation}\label{detdet}
\frac{\det'\Delta_g\;}{\det'\Delta_{g_0}}=e^{-\frac16S_L(g_0,g)},
\end{equation} 
with the usual definition of $\det'\Delta=e^{-\zeta'(0,\Delta)}$ via the spectral zeta function of the laplacian \eqref{zetafunc}. 
Although subleading in $1/k$, the gravitational term in \eqref{logZintr} becomes important in certain situations. When considered on a Riemann surface of genus ${\rm g}\geqslant2$, the gravitational anomaly contributes \cite{KW}, see also \cite{BR1}, a universal finite size correction $\frac16({\rm g}-1)$ to the corresponding geometric adiabatic transport coefficient (anomalous viscosity, introduced in QHE by Avron-Seiler-Zograf \cite{ASZ1}). Also, in the non-singular case, the appearance of regularized spectral determinants \eqref{detdet} was first observed in Ref.\ \cite{ZW}, where large $N$ expansion of the partition function of the 2d Coulomb plasma in planar domains was studied. For the recent work on the gravitational terms in QHE in the $2+1$d Chern-Simons effective theory framework see Refs.\ \cite{AG1,AG2,AG4}, and for the relation between 2d and $2+1$d approaches Ref.\ \cite{KMMW}. 

Another situation, where gravitational term plays a major role is the case of singular Riemann surfaces. These include cones, cusps, and, in the case of Riemann surfaces with boundary, corners. Remarkably, this situation is also experimentally accessible, see Ref.\ \cite{Schine}, where the synthetic Landau levels on a cone were constructed in a photon resonator. Remarkably, geometric defects also occur in graphene \cite{VKG}.

One of the assumptions going into the derivation of \eqref{logZintr} is the $C^\infty(\Sigma)$-smoothness of the metric $g$, when the underlying asymptotic expansion of the Bergman kernel is valid \cite{K}. In the non-smooth case the formula \eqref{logZintr} breaks down, and analysis has to be reconsidered. In this paper we suggest the following formula for $\log Z$ for the integer QH, in the limit of large number of particles
\begin{equation}\label{main}
\log Z=c_2k^2+c_1k+(\zeta(0,\Delta_{\rm sing})-\zeta(0,\Delta_{g_0}))\log (kl_{\rm sing})-\frac12\log \frac{\det'\Delta_{\rm sing}}{\det'\Delta_{g_0}}+c_0+\mathcal O(1/k).
\end{equation}
Here $\zeta(s,\Delta_{\rm sing})$ is the spectral zeta function of the laplacian for the metric with conical singularities on $\Sigma$ and the constants $c_m$ are related to the values of the geometric functionals. The first logarithmic term involves the length scale $l_{\rm sing}$ induced by the singularities, which is roughly the diameter of the singular manifold. This length scale is in general bigger than the magnetic length $l_B=\sqrt{\hbar/eB}$.

We test the formula above for model geometries, where $Z$ can be computed explicitly and find simple expressions for $c_0,c_1,c_2$ as the functions of cone angles. Zeta determinant $\det'\Delta_{\rm sing}$ is a more complicated object, which in our examples is expressed through the Barnes double zeta function \eqref{zetaB}. The study of spectral geometry on manifolds with regular singularities was initiated by Cheeger \cite{Cheeger}, and the formulas for spectral zeta function on cones, relevant to the present work, were obtained in Ref.\ \cite{AS,Dow,SZ,S1}. The standard approach to the spectral zeta function on manifolds with regular singularities involves a combination of heat kernel methods and solving the spectral problem for the appropriate self-adjoint extensions of the associated laplacian \cite{Cheeger,BrS}. 

In this regard Eq.\ \eqref{main} suggests a novel view on the  singular zeta determinants. In order to construct $Z$ on the lhs of \eqref{main} one needs to know only the following ingredients: the holomorphic wave functions on the lowest Landau level (LLL), volume form of the singular metric and the corresponding \kahler potential. Recall that the holomorphic LLL wave functions solve the first order $\bp$-equation on $\Sigma$ and therefore are metric-independent \cite{DK}. Remarkably, this involves no knowledge of the spectrum of $\Delta_g$, -- the relevant information about the spectrum is encoded in the lowest energy level of the magnetic Shr\"odinger equation. 

There is a certain degree of similarity between Eq.\ \eqref{main} and the behavior of the free energy at criticality for the system of characteristic size $l$ in presence of singularities \cite{CP}.  In the singular case the first three terms in \eqref{main}, were recently computed in \cite{LCCW} and the behavior of free energy was interpreted as the emergence of conformal symmetry in the presence of singularities. In another recent work on singularities in QH states \cite{Gr} gravitational anomaly for geometric defects on higher genus curves with $\mathbb Z_n$ symmetry was studied. Landau levels on a cone have been extensively studied in the literature, see e.g.\ \cite{MFBM} and especially \cite{Poux} for the  comprehensive list of references. However in the present work we are only interested in the lowest Landau level, where subtleties associated with the construction of the higher Landau level wave functions do not arise. 

{\bf Acknowledgements.} I would like to thank  T.~Can, A.~Gromov and P. Wiegmann for comments on the manuscript and M. Spreafico for correspondence. This work was partially supported by the German Excellence Initiative at the University of Cologne,  DFG-grant ZI 513/2-1, grant NSh-1500.2014.2 and grant RFBR 15-01-04217.

\section{Integer QH state on curved backgrounds}
\label{back} 

We begin with the definition of the integer QH state on the compact Riemann surface $\Sigma$. Consider a smooth background metric $g_0$ on $\Sigma$, which without loss of generality we choose to be a constant scalar curvature metric, and a constant perpendicular magnetic field $B_0$, with the total flux $k\in\mathbb Z_+$ through the surface $k=\frac1{2\pi}\int B_0\sqrt {g_0}d^2z$. The latter is described by the holomorphic line bundle $L^k$ of degree $k$ on $\Sigma$ with the Hermitian metric $h_0^{k}(z,\bz)$ (magnetic potential), related to the magnetic field as
\begin{equation}\nonumber
B_0=-g_0^{z\bz}\p_z\p_{\bz}\log h_0^{k}(z,\bz).
\end{equation}
The holomorphic part of the one particle wave functions on the lowest Landau level is metric-independent, and is given by holomorphic sections of $L^k$. However, the metric is required to define the $L^2$-orthonormal basis of holomorphic sections $s_j(z),\;j=1,\ldots,N$ according to
\begin{equation}\label{innerprod}
\langle s_j,s_l\rangle|_{g_0}=\frac1{2\pi}\int_\Sigma \bs_j(\bz)s_l(z)h_0^{k}(z,\bz)\sqrt{g_0}d^2z=\delta_{jl}
\end{equation}
The degeneracy $N$ of the LLL is fixed by the Riemann-Roch theorem $N=k+1-\rm g$. The integer QH state for $N$ particles is given by the Slater determinant of the wave functions on completely filled LLL
\begin{equation}\label{iqhest}
\Psi(z_1,...,z_N)=\frac1{\sqrt{N!}}\det [s_j(z_l)]_{j,l=1}^{N},
\end{equation}
and its Hermitian norm is
\begin{equation}\nonumber
|\Psi(z_1,...,z_N)|_{h_0}^2=\frac1{N!}|\det s_j(z_l)|^2\prod_{j=1}^Nh_0^k(z_j,\bz_j).
\end{equation}
By construction, this state is $L^2$ normalized.

Suppose we now have an arbitrary curved metric, which can be conveniently parameterized by the relative \kahler potential $\phi$,
\begin{equation}\label{kahlerpot}
g_{z\bz}=g_{0z\bz}+\p_z\p_{\bz}\phi.
\end{equation}
We would like to define the normalized integer QH state in the curved metric. Setting magnetic field to constant in the new metric 
\begin{equation}\label{constm}
B=-g^{z\bz}\p_z\p_{\bz}\log h^{k}(z,\bz)=k 
\end{equation}
leads to the relation between the magnetic potentials $h^k=h_0^ke^{-k\phi}$. The holomorphic part of the integer QHE state can be taken as in Eq.\ \eqref{iqhest} and its hermitian norm in the new metric is given by
\begin{equation}\nonumber
|\Psi(z_1,...,z_N)|_h^2=\frac1{Z[g_0,g]}\frac1{N!}|\det s_j(z_l)|^2\prod_{j=1}^Nh^k(z_j,\bz_j).
\end{equation}
By construction the normalization factor $Z$ of this state depends on the pair of metrics $g_0,g$. It reads 
\begin{equation}\label{Z}
Z[g_0,g]=\frac1{(2\pi)^NN!}\int_{\Sigma^N}\bigl|\det s_j(z_l)\bigr|^2 \prod_{j=1}^Nh_0^k(z_j,\bz_j)e^{-k\phi(z_j,\bz_j)}\sqrt{g}d^2z_j.
\end{equation}
We can equivalently rewrite $Z$ as the determinant of the matrix of $L^2$ norms
\begin{equation}\label{detform}
Z[g_0,g]=\det \frac1{2\pi} \int_{\Sigma} \bs_j(\bz)s_l(z) h_0^k(z,\bz)e^{-k\phi(z,\bz)}\sqrt gd^2z=\det\langle s_j,s_l\rangle|_{g}.
\end{equation}
In particular, this makes the boundary condition $Z[g_0,g_0]=1$ obvious.
If the basis of sections $s_j$ is orthogonal (but not orthonormal) in the metric $g$, so that $\langle s_j,s_l\rangle|_g=n_j\delta_{jl}$ for some $n_j$, then the generating functional reduces to the product of normalization factors $n_j$ for one particle LLL states
\begin{equation}\nonumber
Z[g_0,g]=\prod_{j=1}^Nn_j.
\end{equation}

The logarithm of $Z$ is called the generating functional. For the smooth metrics $g_0, g$, the generating functional admits large $k$ asymptotic expansion with the coefficients given by geometric functionals of the metrics. The first few terms of the expansion read \cite[Eq.\ 5.2]{K},
\begin{align}\nonumber
\log Z=&-k^2 S_{AY}(g_0,\phi)+\frac k2S_M(g_0,\phi)+\frac16S_L(g_0,\phi)\\\label{logZ}
&-\frac5{192k}\left(\frac1{2\pi}\int_\Sigma R^2\sqrt gd^2z-\frac1{2\pi}\int_\Sigma R_0^2\sqrt{g_0}d^2z\right)+\mathcal O(1/k^2).
\end{align}
Here the subleading terms in $1/k$ consist of integrals of curvature invariants of the metric, i.e., given by polynomials in scalar curvature $R$ and its derivatives. The first three terms are the given by the Aubin-Yau, Mabuchi and Liouville functionals of the \kahler potential
\begin{align}\label{SAY}
&S_{AY}(g_0,\phi)=\frac1{2\pi}\int_\Sigma\bigl(\phi\,\p_z\p_{\bz}\phi+\phi\sqrt{g_0}\bigr)\,d^2z,\\
\label{SM}
&S_M(g_0,\phi)=\frac1{2\pi}\int_\Sigma\left(-\frac12\phi R_0\sqrt{g_0}+\sqrt g\log\frac{\sqrt g}{\sqrt{g_0}}\right)d^2z,\\
\label{SL}
&S_L(g_0,\phi)=\frac1{2\pi}\int_\Sigma\left(-\log\frac{\sqrt g}{\sqrt{g_0}}\p_z\p_{\bz}\log\frac{\sqrt g}{\sqrt{g_0}}+\frac12R_0\sqrt{g_0}\log\frac{\sqrt g}{\sqrt{g_0}}\right)d^2z.
\end{align}
The important assumption in the derivation of the expansion Eq.\ \eqref{logZ} is smoothness of the metrics. In particular, we can immediately see, that the $\int R^2$ term blows up when curvature has delta-function singularities. 

Now we consider what happens when the metric $g$ is singular.

\section{Conical singularities and geometric functionals}

Two types of singularities are of particular interest in the quantum Hall states. One is the well-known magnetic singularities, when the magnetic field has the form $B(z)=B_0(z)-2\pi \varphi \delta(z_0)$ with $B_0(z)$ being a smooth function and $\delta(z_0)$ is the Dirac delta function at the point $z_0$. In the integer quantum Hall effect the delta-function term with $\varphi\in \mathbb R$ corresponds to the Aharonov-Bohm flux, when the surface is not simply-connected as e.g. in the Laughlin's quantization argument \cite{L1}. In the fractional quantum Hall states the delta-function singularity creates a quasi-hole excitation \cite{L} with the charge $\varphi\in \mathbb Z_+$ quantized on a compact surface. 

A more rigid type of singularity is the curvature singularity of the metric on the surface. In the case of conical singularities at points $z_\alpha$ the Ricci tensor 
\begin{equation}\label{ricci}
R_{z\bz}=-\p_z\p_{\bz}\log g_{z\bz}
\end{equation}
 for the metric $ds^2=2g_{z\bz}dzd\bz$ on $\Sigma$ in local complex coordinates has the form 
\begin{equation}\label{curv}
R_{z\bz}=2\lambda g_{z\bz}+\pi\sum_{\alpha}(1-a_\alpha)\delta^2(z_\alpha),
\end{equation}
where $\delta^2(z_\alpha)$ is the Dirac delta-function, which in a local euclidean chart $U_\alpha$ around $z_\alpha$ satisfies $\int_{U_\alpha}\delta^2(z_\alpha)d^2z=1$. The conical singularities are characterized locally by the opening angle $2\pi a_\alpha$ with $0\leq a_{\alpha}<1$, where $a_{\alpha}\to0$ limit corresponds to a cusp and $a\to1$ limit to a smooth surface. In Eq.\ \eqref{curv} we also implicitly assumed that it is possible to choose a metric of constant scalar curvature outside the singular points, which will be the case in specific examples. The curvature singularities can be considered both in the integer and fractional QH states, but here we focus on the integer QH state. 
Since we are interested in the purely gravitational contribution to the generating functional, we set the AB-fluxes to zero $\varphi=0$, although we shall note that both types of singularities can be considered simultaneously \cite{LCCW}. 

Near the conical point $z_\alpha$ the metric has the following singular behavior
\begin{equation}\nonumber
g_{z\bz}\sim|z-z_\alpha|^{-2(1-a_\alpha)}f_{\rm reg}(z),
\end{equation}
as follows from Eqns.\ \eqref{ricci} and \eqref{curv} and the identity $\p_z\p_{\bz}\log|z-z_\alpha|^2=\pi\delta^2(z_\alpha)$.
Hence, for a smooth background metric $g_0$ the \kahler potential in \eqref{kahlerpot} is smooth at the singular point. Then it is not hard to check that the Aubin-Yau and Mabuchi functionals (\ref{SAY},\ref{SM}) are finite since all the terms are integrable. In the Liouville functional only the second term is integrable and the kinetic term behaves as $\delta^2(z_\alpha)\log|z-z_\alpha|$, hence it diverges. The higher-order terms in the expansion of $\log Z$ also diverge since they contain integrals of products of delta-functions at conical points. Hence the smooth-case expansion \eqref{logZ} breaks down at the order $O(1)$.

In the next two sections we will compute asymptotics of $\log Z$ for two model singular geometries and check the proposed general form \eqref{main} of the expansion in the singular case.

\section{Integer QH state on the spindle}

The spindle (or, american football) $\Sigma_a$, see Fig.\ 1 (left), is the sphere with two antipodal conical points with equal opening angles $2\pi a,\,0<a\leq1$ with $a=1$ corresponding to the round sphere. The exists a metric of constant scalar curvature outside the conical points on this geometry \cite{T} which reads
\begin{equation}\label{spindlemetric}
ds^2=2g_{z\bz}dzd\bz=\frac{2adzd\bz}{|z|^{2(1-a)}(1+|z|^{2a})^2}.
\end{equation}
We normalized the total area as $\int_{\Sigma_a}\sqrt gd^2z=2\pi$ and in our notations $\sqrt g=2g_{z\bz}$. The Ricci curvature of this metric is given by
\begin{equation}\nonumber
R_{z\bz}=2ag_{z\bz}+\pi(1-a)\delta^2(0)+\pi(1-a)\delta^2(\infty),
\end{equation}
and the Euler formula holds $\int_{\Sigma_a}R\sqrt gd^2z=8\pi$ since the surface is topologically a sphere. At $a=1$ the metric \eqref{spindlemetric} reduces to the standard round metric on the sphere  $g_{0z\bz}=\frac{1}{(1+|z|^{2})^2}$. Thus the \kahler potential \eqref{kahlerpot} for the spindle relative to the round metric reads
\begin{equation}\label{spindle}
\phi_{\rm sp}=\log\frac{(1+|z|^{2a})^{1/a}}{1+|z|^2}.
\end{equation}
The constant magnetic field $B_0=k$ on the round sphere and on the spindle $B=k$ correspond to the following magnetic potentials 
\begin{align}\nonumber
h_0^k(z,\bz)=\frac1{(1+|z|^{2})^{k}},\quad h^k(z,\bz)=h_0^ke^{-k\phi_{\rm sp}}=\frac1{(1+|z|^{2a})^{k/a}}.
\end{align}

\begin{figure}[t]
\begin{center}
\begin{tikzpicture}
    \draw[scale=2,domain=-1:1,smooth,variable=\x,purple,very thick]  plot ({-(-(\x)^2+3)^.5},{\x});
        \draw[scale=2,domain=-1:1,smooth,variable=\x,purple,very thick]  plot ({(-(\x)^2+3)^.5-2.835},{\x});
                \draw[scale=1,domain=-.61:.63,dashed,variable=\x,purple,very thick]  plot ({\x-2.8},{(-(\x)^2+3)^.5-1.6});
                                \draw[scale=1,domain=-.65:.61,smooth,variable=\x,purple,very thick]  plot ({\x-2.8},{-(-(\x)^2+3)^.5+1.6});
                                    \draw[black] (-2.84,1.9) node[below]{$a$};
                                                                        \draw[black] (-2.84,-1.4) node[below]{$a$};
    \draw[scale=1,domain=-1/2:1/2,smooth,variable=\x,blue,very thick]  plot ({\x-.988+2},{2.5*\x});
        \draw[scale=1,domain=-1/2:1/2,smooth,variable=\x,blue,very thick]  plot ({-\x+2},{2.5*\x});
                \draw[scale=1.1,domain=-.9:.9,dashed,variable=\x,blue,very thick]  plot ({\x-.405+1.8},{(-(\x)^2+3)^.5-2.6});
                                \draw[scale=1,domain=-1:1,smooth,variable=\x,blue,very thick]  plot ({\x-.499+2},{-(-(\x)^2+3)^.5+.178});
                                    \draw[black] (-.5+2,1) node[below]{$a$};
  \end{tikzpicture}
  \vspace{0.5cm} \\
{\small Figure 1.\, Spindle $\Sigma_a$ ("american football") vs. simple cone $D_a$.}
\end{center}
\end{figure}

On the round sphere, the orthonormal basis of sections with respect to the inner product \eqref{innerprod} reads 
\begin{align}\label{basis}
s_j(z)=\sqrt{\frac{(k+1)!}{(k-j+1)!(j-1)!}}\;z^{j-1},\quad j=1,\ldots,k+1.
\end{align}
Plugging the basis \eqref{basis} into the determinantal formula \eqref{detform} for the generating functional and performing the integrals we obtain
\begin{align}\label{genf}\nonumber
&\log Z_{\rm sp}=\log\det M_{ij},\\
&M_{ij}=\frac{(k+1)!}{(k-j+1)!(j-1)!}\,B\left(1+\frac{j-1}a,1+\frac{k+1-j}a\right)\cdot\delta_{ij},
\end{align}
where $B(x,y)$ is the standard Beta-function. Thus we arrive at the following formula for the normalization factor
\begin{align}\label{Zsp}
Z_{\rm sp}=\left(\frac{\Gamma(2+k)}{\Gamma\bigl(2+\frac ka\bigr)}\right)^{k+1}\cdot\left(\frac{\prod_{j=1}^{k}\Gamma\bigl(1+ \frac ja\bigr)}{G(2+k)}\right)^2,
\end{align}
where $G(z)$ is the Barnes $G$-function, see e.g. \cite{GR}. As a consistency check we immediately see that the boundary condition $Z|_{a=1}=1$ holds, since $G(n)=\prod_{j=1}^{n-2}\Gamma(j+1)$ for integer $n$.

The large $k$ asymptotics of $\log Z$ can be now evaluated using Euler-Maclaurin summation formula, see Appendix A for details,
\begin{align}\nonumber
\log Z_{\rm sp}=&-\frac{1-a}{2a}k^2+\frac{\log a}2 k+\frac16\left(a+\frac1a-2\right)\log k\\\label{logZ1}&
+2\zeta'_2\bigl(0,a,1,1\bigr)-2\zeta_R'(-1)+\frac{13}{12}(1-a)+\frac32\log a+\mathcal O(1/k)
\end{align}
where $\zeta_2$ is the Barnes double zeta function, defined by the following double series 
\begin{equation}\label{zetaB}
\zeta_2(s,a,b,x)=\sum_{m,n=0}\frac1{(am+bn+x)^s}.
\end{equation}
For more details on this function we refer to \cite{S2} and references therein.

Note that non-local $k^2\log k$ and $k\log k$ terms cancel in \eqref{logZ1}, which is an important consistency check. The $\log k$ and $\mathcal O(1)$ terms are related to the spectral zeta function for the scalar laplacian $\Delta_g$, defined as 
\begin{equation}\label{zetafunc}
\zeta(s,\Delta_g)=\sum_{\lambda\neq 0}\lambda^{-s},
\end{equation}
where the sum goes over all positive eigenvalues $\lambda$ with multiplicities. The zeta determinant of laplacian on the spindle with cone angle $2\pi a$ was computed in \cite{SZ}, see Appendix B for more details. Here we quote the result in Eq.\ \eqref{zetaprime} 
\begin{equation}\nonumber
\log{\det}'\Delta_g=-\zeta'(0,\Delta_g)=-4\zeta_2'(0,a,1,1)+\frac a2-2\log a-\left(\frac a6+\frac1{6a}-1\right)\log\frac a2.
\end{equation}
At $a=1$ this reduces to the value of the zeta determinant on the round sphere
\begin{align}
\log{\det}' \Delta_{g_0}=-4\zeta'(-1)+\frac12-\frac23\log2,
\end{align} 
with the area normalized to $2\pi$, see Eq.\ \eqref{zetaround1}. We also need the value of the zeta function at zero: $\zeta(0,\Delta_g)=-1+\frac a6+\frac1{6a}$ and $\zeta(0,\Delta_{g_0})=-\frac23$. Then we can recast the formula for the generating functional \eqref{logZ1} in the form announced in Eq.\ \eqref{main}
\begin{align}\label{logZsp}
\log Z_{\rm sp}=c_2k^2+c_1k+\bigl(\zeta(0,\Delta_g)-\zeta(0,\Delta_{g_0}\bigr)\log kl_{\rm sing}-\frac12\log\frac{{\det}' \Delta_{g}}{{\det}'\Delta_{g_0}}+c_0+\mathcal O(1/k)
\end{align}
It is straightforward to check that the first two terms in \eqref{logZsp} coincide with the values of the Aubin-Yau and Mabuchi functionals, computed for the \kahler potential $\phi_{\rm sp}$ and the round metric, 
\begin{align}\label{AY}
c_2=-S_{AY}(g_0,\phi_{\rm sp})=-\frac{1-a}{2a},\quad
c_1=\frac12S_M(g_0,\phi_{\rm sp})=\frac12\log a,
\end{align}
and thus to the order $\mathcal O(k)$ the expansion if consistent with the smooth case \eqref{logZ}. There is also a correction constant $c_0$, which has a rather simple form
\begin{equation}
c_0=\frac56(1-a+\log a).
\end{equation}
We have not been able to establish the exact origin of this constant. The natural guess is that it is presumably related to the value of the Liouville functional in this geometry, which needs to be appropriately regularized. Finally, the length scale here is given by $l_{\rm sing}=\sqrt{2/a}$ and it is related to the distance between the two conical point as $d(0,\infty)=\pi l_{\rm sing}/2$.

\section{IQHE on a cone}

Here we make the same calculation on the simple cone $D_a$ (see Fig. 1, right) with the angle $2\pi a$ and flat metric outside the singularity. We parameterize $D_a$ by the complex coordinate $z$ in the unit disk $0\leqslant |z|\leqslant1$. The metric on the cone
\begin{equation}\label{conemet}
ds^2=\frac{2adzd\bz}{|z|^{2(1-a)}}
\end{equation}
and has the area $2\pi$. Then we can solve the condition \eqref{constm} of the constant perpendicular magnetic field and find the magnetic potential 
\begin{equation}\nonumber
h^k(z,\bz)=e^{-\frac{k}{a}|z|^{2a}}.
\end{equation}
The flux of the magnetic field, when restricted to $D_a$ equals $\int_{D_a}B\sqrt gd^2z=2\pi k$.
On the flat disk $a=1$ the orthonormal basis of holomorphic LLL states is given by
\begin{equation}\nonumber
s_j(z)=\sqrt{\frac{k^j}{(j-1)!}}z^{j-1},\quad j\leqslant k
\end{equation}
This basis is infinite when considered on $\mathbb C$, therefore we introduce the cutoff $j\leqslant k$, restricting to the first $k$ states supported inside the disk. 

Now we use this basis to determine the normalization factor \eqref{detform} of the integer QH state one the cone $D_a$,
\begin{equation}\label{Zcone}
Z_{\rm cone}=\prod_{j=1}^kk^j\left(\frac ka\right)^{-\frac{j-1}a-1}\frac{\Gamma\bigl(\frac{j-1}a+1\bigr)}{\Gamma(j)},
\end{equation}
The asymptotic expansion of $\log Z_{\rm cone}$ at large $k$ is obtained in Eq.\ \eqref{Zconeas} in Appendix A and explicit formulas for the spectral zeta function on the cone with Dirichlet boundary conditions are presented in Eqns.\ (\ref{zetac1}-\ref{zetac3}) in Appendix B. Putting these together we arrive at the relation  
\begin{align}\label{logZ22}
\log Z_{\rm cone}=b_0k^2+b_1k+\bigl(\zeta(0,\Delta_g)-\zeta(0,\Delta_{g_0}\bigr)\log kl_{\rm sing}-\frac12\log\frac{{\det}' \Delta_{g}}{{\det}'\Delta_{g_0}}+b_2+\mathcal O(1/k),
\end{align}
where $\zeta(0,\Delta_g)=\frac1{12}(a+1/a)$ and $\zeta(0,\Delta_{g_0})=\frac16$. The constants in this case are given by
\begin{align}\nonumber
&b_0=-\frac{3(1-a)}{4a},\quad b_1=\frac12\left(\frac1a-1+\log a\right),\\\nonumber
&b_2=\frac5{24}(1-a)+\frac16\log a.
\end{align}
The length scale $l_{\rm sing}=\sqrt{2/a}$ in this case is the distance between the conical point and the edge of the cone, $d(0,1)=l_{\rm sing}$.

\section{Discussion}
In this note we suggest and check in simple examples the following formula for the normalization factor of the integer QH state on singular Riemann surfaces
\begin{equation}\label{main1}
Z=\left[\frac{\det'\Delta_{\rm sing}}{\det'\Delta_{g_0}}\right]^{-1/2}\cdot (kl_{\rm sing})^{\zeta_{\Delta_{\rm sing}}(0)-\zeta_{\Delta_{g_0}}(0)}\cdot e^{c_2k^2+c_1k+c_0+\mathcal O(1/k)}.
\end{equation}
Here $g_0$ is a smooth background metric, $\det'\Delta_{\rm sing}$ is the zeta regularized spectral determinant of the laplacian on a singular surface, $\zeta_{\Delta_{\rm sing}}(s)$ is the spectral zeta function, $l_{\rm sing}$ is a certain length scale associated with singularities and $c_m$ are simple constants related to the values of geometric functionals. In this work we tested this formula for the model geometries of the flat cone and the spindle, Fig.\ 1. 

Some aspects of the relation Eq. \eqref{main1} remain to be better understood. In particular,
note that  in view of Eq.\ \eqref{detdet}, the power of $\det'\Delta_{\rm sing}$ in Eq.\ \eqref{main1} is half of its smooth case value \eqref{logZ}. One the one hand this is not completely surprising, since the expansion of $\log Z$ breaks down at the order $\mathcal O(1)$ where $S_L$ starts to diverge, and there is no aprori reason to expect that Eq.\ \eqref{logZ} is valid in the singular case. However, it is possible that Eq.\ \eqref{logZ} continues to hold in come regularized sense, and the value of the constant $c_0$ could be equal to the $\frac1{12}S_L^{\rm reg}$ for the regularized Liouville functional on the singular geometry. Also the definition of $l_{\rm sing}$ remains to be understood for more general geometries. 

Another interesting aspect of Eq.\ \eqref{main1} that we would like to discuss is the appearance 
of the Barnes double zeta function in the explicit formulas \eqref{logZ2}, \eqref{Zconeas} for $Z$.    It would be interesting to compute $O(1)$ term in the norm of the Laughlin state on singular surface characterized by the metric $g_{\rm sing}$. In the notations of Eq.\ \eqref{Z} this is given by the following integral
\begin{align}\nonumber
Z_\beta[g_0,g_{\rm sing}]=\frac1{N_0}\int_{\mathbb C^{k+1}}|\det s_j(z_m)|^{2\beta}\prod_{j=1}^{k+1}h^{\beta k}(z_j,\bz_j)\sqrt gd^2z_j,
\end{align}
where $\beta\geqslant1$ and $N_0$ is a normalisation constant chosen such that in the smooth limit $g_{\rm sing}\to g_0$, we have $Z_\beta[g_0,g_0]=1$. 

It is reasonable to expect that the terms up to $O(k)$ in $\log Z$ should coincide with the smooth case \cite[Eq.\ 1.1]{FK}, i.e., are given by the values of the geometric functionals computed on the singular geometry. Conversely, we expect the  $O(1)$ term to be a non-trivial $\beta$-deformation of the formulas for zeta determinants (e.g., one obvious guess is that $\zeta_2'(0,a,b,c)$ will appear with values of arguments $b,c\neq1$). 

In Ref.\ \cite{LCCW} the large $k$ behavior of the generating functional \eqref{main} was interpreted as emergence of conformal symmetry in presence of singularities. Better understanding of the $O(1)$ term in Eq.\ \eqref{main} could help answer the question: exactly what kind of conformal theory emerges in the singular case? If one is to think of the conical singularities as insertions vertex operators of charge $a$, then after the subtraction of large magnetic field terms, the $O(1)$ term could be interpreted as a correlation function in the appropriate CFT. There is an example of a CFT where Barnes special functions, such as Barnes double gamma function (closely related to $\zeta_2'$), show up in correlation functions, -- namely, the Liouville field theory \cite{DO,ZZ}. Emergence of the Liouville quantum gravity in QH states at the tip of a cone is a thrilling possibility\footnote{Relevance of the Liouville theory to the description of QH states on singular surfaces has also been advocated by P. Wiegmann.}, which we suggest as an avenue for further investigations.

\section*{Appendix A}

Here we derive the asymptotics of $\log Z$ (\ref{logZ1},\ref{logZ22}). We begin we a collection of useful asymptotic formulas, which can be found e.g. in Ref.\ \cite{GR}. At large $z$ the Gamma function, Barnes $G$-function, digamma function $\psi(z)$ and polygamma function $\psi_n(z)$ admit the following expansions 
\begin{align}\label{G1}
&\log \Gamma(1+z)=\left(z+\frac12\right)\log z-z+\frac12\log 2\pi+\frac{1}{12z}+\mathcal O(1/z),\\
\label{G3}
&\log G(1+z)=\left(\frac{z^2}2-\frac1{12}\right)\log z-\frac34z^2+\frac z2\log 2\pi+\zeta_{R}'(-1)+\mathcal O(1/z),\\
\label{G5}
&\psi(1+z)=\frac{\Gamma'(1+z)}{\Gamma(1+z)}=\log z+\mathcal O(1/z),\\\label{G6}
&\psi_n(1+z)=\frac{d^n\psi(1+z)}{dz^n}=\mathcal O(1/z),\quad n>1
\end{align}
where $\zeta_R(s)$ is the Riemann zeta function and $\zeta_{R}'(s)=\frac{d}{ds}\zeta_R(s)$. We will need the values of the digamma and polygamma functions at $1$:
\begin{align}\nonumber
&\psi(1)=-\gamma,\\
&\psi_{n}(1)=(-1)^{n+1}n!\,\zeta_R(n+1),\quad n\geqslant1,
\end{align}
where $\gamma$ is the Euler-Mascheroni constant. More generally
\begin{align}\nonumber
\psi_{n}(z)=(-1)^{n+1}n!\,\zeta_H(n+1,z),
\end{align}
where the Hurwitz zeta function is defined as
\begin{align}\nonumber
\zeta_H(s,z)=\sum_{j=0}^\infty\frac1{(j+z)^s},
\end{align}
see e.g. \cite{GR}.
The first derivative of the Barnes double zeta function $\zeta'_2(s,a,b,x)=\frac{d}{ds}\zeta_2(s,a,b,x)$ admits the following asymptotic expansions \cite{S2} for large and small $a$, respectively,
\begin{align}\nonumber
a\gg1,\quad\zeta'_2(0,a,1,1)=&-\frac1{12}\left(3+a+\frac1a\right)\log a+\left(\frac1{12}-\zeta'_R(-1)\right)a\\&-\frac14\log 2\pi+\frac{\gamma}{12a}+\sum_{j=2}^\infty\frac{B_{2j}\,\zeta_R(2j-1)}{2j(2j-1)}a^{-(2j-1)}\\\nonumber
a\ll1,\quad\zeta'_2(0,a,1,1)=&\left(\frac1{12}-\zeta'_R(-1)\right)\frac1a-\frac14\log 2\pi+\frac{\gamma a}{12}+\sum_{j=2}^\infty\frac{B_{2j}\,\zeta_R(2j-1)}{2j(2j-1)}a^{2j-1}
\end{align}
The following integral appears below
\begin{align}\label{intgamma}
\int_0^z\log\Gamma\left(1+x\right)dx=-\frac12z(z+1)+\frac12z\log 2\pi+z\log \Gamma(1+z)-\log G(1+z).
\end{align}
The large $k$ asymptotics of the sum of $\Gamma$-functions can be calculated via the Euler-Maclaurin formula
\begin{align}\nonumber
\sum_{j=1}^k\log\Gamma\left(1+\frac ja\right)
=&\int_0^k\log\Gamma\left(1+\frac xa\right)dx+\frac12\log\Gamma\left(1+\frac ka\right)+\frac1{12a}\left(\psi\left(1+\frac ka\right)-\psi(1)\right)\\\nonumber
&+\sum_{j=2}^\infty\frac{B_{2j}}{(2j)!}\left(\psi_{2(j-1)}\left(1+\frac xa\right)\big|_{x=k}-\psi_{2(j-1)}\left(1+\frac xa\right)\big|_{x=0}\right).
\end{align}
Now we use Eq.\ \eqref{intgamma} and formulas (\ref{G1}, \ref{G3}, \ref{G5}, \ref{G6}) to find the large $k$ asymptotics
\begin{align}\nonumber
\sum_{j=1}^k\log\,\Gamma\left(1+\frac ja\right)
=&\,\frac1{2a}k^2\log k-\left(\frac{1}{2a}\log a+\frac3{4a}\right)k^2+\left(\frac1{2a} +\frac12\right)k\log k\\\nonumber
&-\left(\frac12\log a+\frac1{2a}\log a+\frac12+\frac1{2a}-\frac12\log 2\pi\right)k+\frac1{12}\left(3+a+\frac1{a}\right)\log k\\
&+\zeta'_2(0,a,1,1)+\frac12\log 2 \pi+\mathcal O(1/k).
\label{Gprod}
\end{align}
Another way to compute this asymptotic expansion is to use the exact formula 
\begin{equation}\label{exactfor}
\prod_{j=1}^k\Gamma\left(1+\frac ja\right)=k!(2\pi)^{\frac{k+1}2}a^{-\frac{k^2}{2a}-\bigl(1+\frac1a\bigr)\frac k2-\frac12}\bigl[\Gamma_2(k+1,1,a)\bigr]^{-1}.
\end{equation}
This follows by induction from the functional equation \cite[Prop.\ 8.6]{S2} for the Barnes double Gamma function $\Gamma_2$ (see e.g., \cite[Sec.\ 1]{S2} for the definition), taking into account $\Gamma_2(1,a,1)=\sqrt{2\pi}a^{-1/2}$. Then asymptotics \eqref{Gprod} follows from the asymptotic expansion of the logarithm of $\Gamma_2$ \cite[Prop. 8.11]{S2}. 

Now we use the asymptotic formula \eqref{Gprod} and Eqns.\ (\ref{G1}, \ref{G3}) in order to find asymptotics of the generating functionals Eqns.\ (\ref{Zsp}, \ref{Zcone})
\begin{align}\nonumber
\log Z_{\rm sp}=&\frac{a-1}{2a}k^2+\frac{\log a}2 k+\frac16\left(a+\frac1a-2\right)\log k+2\zeta'_2(0,a,1,1)-2\zeta_R'(-1)\\\label{logZ2}&+\frac{13}{12}(1-a)+\frac32\log a+\mathcal O(1/k),\\\nonumber
\log Z_{\rm cone}=&-\frac{3(1-a)}{4a}k^2+\left(\frac1a-1+\log a\right)\frac k2+\left(\frac a{12}+\frac1{12a}-\frac16\right)\log k\\\label{Zconeas}&+\zeta'_2\bigl(0,a,1,1\bigr)-\zeta_R'(-1)+\frac12\log a+\mathcal O(1/k)
\end{align}
It is not hard to show that 
\begin{align}\label{zetaRprime}
\zeta'_2(0,1,1,1)=\zeta_{R}'(-1),
\end{align}
hence the boundary condition $Z|_{a=1}=1$ holds.

\section*{Appendix B}

Here we collect formulas for the spectral zeta function on the conical surfaces, derived in Refs.\ \cite{SZ,S1,S2}. In order to compare the results with Eqns.\ \eqref{logZ2}, \eqref{Zconeas}, we  express the answers in terms of $\zeta'_2\bigl(0,a,1,1\bigr)$ and also adjust the formulas, taking into account the normalization of the area $A=2\pi$, which is adopted in this paper for consistency with previous work. 

In Ref.\ \cite{SZ} the spectral zeta function is computed for the metric on the spindle
\begin{equation}\label{SZmet}
d^2s=d\theta^2+\frac1{a^2}\sin^2\theta d\varphi^2.
\end{equation}
The cone angle here is $2\pi/a$ and  the  area of the surface in this metric equals $4\pi/a$. Under the coordinate change $\tan\theta/2=1/r^a$ and $z=re^{i\varphi}$ the metric above reads
\begin{equation}\nonumber
d^2s=\frac{4dzd\bz}{a^2|z|^{2(1-1/a)}(1+|z|^{2/a})^2}.
\end{equation}

We start with the formula \cite[Thm.\ 4.16]{SZ} for the spectral zeta function\footnote{Note the corrected \cite{Scomm} coefficient in the fifth term, cf. \cite[Thm.\ 4.16]{SZ}.} in the metric \eqref{SZmet}
\begin{align}\nonumber
&\zeta'\bigl(0,\Delta_{S^2_{1/a}}\bigr)=-\left(\frac a3+\frac1{3a}\right)\log a -2\log 2\pi +\frac a3+1+ \frac1{2a}+\log \Gamma\left(1+\frac1a\right)\\\label{zeta3}
&-2a\zeta_R'(-1)-2a\zeta_H'\biggl(-1,1+\frac1a\biggr)+2i\int_0^\infty\log\frac{\Gamma(1+i\frac ya)\Gamma(1+\frac1a+i\frac ya)}{\Gamma(1-i\frac ya)\Gamma(1+\frac1a-i\frac ya)}\frac{dy}{e^{2\pi y}-1}.
\end{align}
Now, we use the following formula \cite[Prop. 5.1]{S2} for the first derivative of the Barnes double zeta function at $s=0$
\begin{align}\nonumber
&\zeta_2'(0;a,b,x)=\left(-\frac12\zeta_H\left(0,\frac xa\right)+\frac ab\zeta_H\left(-1,\frac xa\right)-\frac1{12}\frac ba\right)\log a+\frac12\log \Gamma\left(\frac xa\right)\\\nonumber&-\frac14\log 2\pi-\frac ab\zeta_H\left(-1,\frac xa\right)-\frac ab\zeta_H'\left(-1,\frac xa\right)+i\int_0^\infty\log\frac{\Gamma(\frac{x+iby}a)}{\Gamma(\frac{x-iby}a)}\frac{dy}{e^{2\pi y}-1}
\end{align}
in order to express the last integral in \eqref{zeta3} in terms of $\zeta_2'(0;a,b,x)$ as follows
\begin{align}\nonumber
\zeta'\bigl(0,\Delta_{S^2_{1/a}}\bigr)&= 2\zeta_2'(0,a,1,a)+2\zeta_2'(0,a,1,a+1)-\frac1{2a}-\log 2\pi\\
&=4\zeta_2'(0,a,1,1) -\frac1{2a}+ \log a.\label{secline}
\end{align}
The relevant values of the Hurwitz zeta function can be found, e.g., in \cite{GR,S2},
\begin{align}\nonumber
\zeta_H(-1,z)=-\frac12z^2+\frac12z-\frac1{12},\quad
\zeta_H(0,z)=\frac12-z,\quad\zeta_H(s,1)=\zeta_R(s).
\end{align}
In the second line in Eq.\ \eqref{secline} we use the formulas
\begin{align}\nonumber
&\zeta_2'(0,a,1,a)=\zeta_2'(0,a,1,1)+\frac12\log a\\\nonumber
&\zeta_2'(0,a,1,a+1)=\zeta_2'(0,a,1,1)+\frac12\log 2\pi,
\end{align}
which easily follow from the definition \eqref{zetaB}. The following relation 
\begin{align}\nonumber
\zeta_2'(0,1/a,1,1)=\zeta_2'(0,a,1,1)+\left(\frac14+\frac{a}{12}+\frac1{12a}\right)\log a.
\end{align}
also follows from \eqref{zetaB}. This relation is needed to invert the cone angle $1/a\to a$, in order to compare with the metric \eqref{spindlemetric}. We obtain 
\begin{align}
&\zeta'\bigl(0,\Delta_{S^2_{a}}\bigr)=4\zeta_2'(0,a,1,1)-\frac{a}2+\frac13\left(a+\frac1{a}\right)\log a,
\end{align}
for the spindle with cone angle $a$ and the metric
\begin{equation}\label{samet}
d^2s=\frac{4a^2dzd\bz}{|z|^{2(1-a)}(1+|z|^{2a})^2}.
\end{equation}
This metric has the area $4\pi a$ and differs from the metric \eqref{spindlemetric}, which has the area $2\pi$, by a factor $2a$. Hence the corresponding laplacians are related as $\Delta_g=2a\Delta_{S^2_a}$, where $\Delta_g=2g^{z\bz}\p_z\p_{\bz}$ is the laplacian in the metric \eqref{spindlemetric} and $\Delta_{S^2_a}$ is the laplacian in the metric \eqref{samet}. Using the relation 
\begin{equation}\label{zetaC}
\zeta'(0,C\Delta_g)=\zeta'(0,\Delta_g)-\zeta(0,\Delta_g)\log C, 
\end{equation}
we obtain 
\begin{equation}\nonumber
\zeta'(0,\Delta_g)=\zeta'\bigl(0,\Delta_{S^2_{a}}\bigr)-\zeta(0,\Delta_g)\log 2a.
\end{equation}
The value of the zeta function at zero can be read off from Ref. \cite[Thm.\ 4.15]{SZ}
\begin{equation}\label{zeta0}
\zeta(0,\Delta_g)=-1+\frac a6+\frac1{6a}.
\end{equation}
Thus we arrive at the final formula for the logarithm of the regularized determinant of laplacian in the metric \eqref{spindlemetric},
\begin{equation}\label{zetaprime}
\zeta'(0,\Delta_g)=4\zeta_2'(0,a,1,1)-\frac a2+2\log a+\left(-1+\frac a6+\frac1{6a}\right)\log\frac a2.
\end{equation}
In the smooth limit at $a=1$ the values of zeta function Eqns.\ \eqref{zeta0} and \eqref{zetaprime} reduce to the following values
\begin{align}\label{zetaround}
&\zeta(0,\Delta_{g_0})=-\frac23,\\\label{zetaround1}
&\zeta'(0,\Delta_{g_0})=4\zeta'_R(-1)-\frac12+\frac23\log 2,
\end{align}
where we used \eqref{zetaRprime}. The standard value of $\zeta'(0)$ on the sphere of area $4\pi$ is $\zeta'(0,\Delta_{S^2})=4\zeta'_R(-1)-\frac12$ and the $\log 2$ term in \eqref{zetaround1} is because the area of the sphere in the metric $g_0$ is $2\pi$.

Explicit formulas for the zeta function on the cone were obtained in \cite{S1}. In the notations of \cite{S1} the cone has the area $\pi l^2/\nu$, where $\nu=1/a$. Again, using the relation \eqref{zetaC} we can adopt the formulas in \cite{S1} to our case
\begin{align}\label{zetac1}
&\zeta(0,\Delta_{g_0})=\frac16,\quad\zeta(0,\Delta_{g})=\frac1{12}\left(a+\frac1a\right),\\\label{zetac2}
&\zeta'(0,\Delta_{g_0})=2\zeta'_R(-1)-\frac16\log2+\frac5{12}+\frac12\log2\pi,\\\label{zetac3}
&\zeta'(0,\Delta_g)=2\zeta'_2(0,a,1,1)+\frac1{12}\left(a+\frac1a\right)\log\frac a2+\frac{5a}{12}+\frac12\log a+\frac12\log2\pi,
\end{align}
where the metric $g$ on the cone is given in Eq.\ \eqref{conemet}, $g_0$ is the metric on the disk $a=1$, and both metrics have the area $2\pi$.


\end{document}